\documentclass[preprint,prd,aps,pdftex,showpacs,preprintnumbers,amsmath,amssymb]{revtex4-1}
\usepackage{graphics,epsfig,subfigure}
\usepackage{diagbox}
\usepackage[usenames]{color}
\usepackage{color}
\usepackage{graphicx}
\usepackage{amsfonts}
\usepackage{indentfirst}
\usepackage{booktabs}
\usepackage{hyperref}
\hypersetup{hypertex=ture,backref=true,colorlinks=ture,linkcolor=blue,anchorcolor=blue,citecolor=blue}
\usepackage{float}
\usepackage{multirow}
\usepackage{tensor}

\begin{document}

\title{Frozen gravitational stars in Einsteinian cubic gravity}

\author{Yong-Qiang Wang}\email{yqwang@lzu.edu.cn, corresponding author}
\affiliation{School of Physical Science and Technology, Lanzhou University, Lanzhou 730000, China}

\begin{abstract}
In this paper, we investigate the numerical solutions for spherically symmetric situations in Einstein cubic gravity. In addition to the previously found black hole solutions, we uncover a new class of solutions that lack horizons. Due to the divergence of the central curvature, these solutions represent a novel type of naked singularity. By varying parameters, we find that under certain conditions, the metric function can approach zero infinitely, indicating the emergence of a critical event horizon. These solutions, featuring a critical horizon and obtained solely from pure gravity model, can be described as frozen gravitational stars (FGSs). Additionally, the radius of this horizon is linked to the coupling constant lambda, increasing as lambda increases. Remarkably, from the perspective of an external observer, these solutions closely resemble extreme black holes. Notably, we find that for all values of lambda, the position of this critical horizon coincides with that of a Schwarzschild black hole of the same mass.
\end{abstract}

%\pacs{11.10.Kk., 04.50.+h.\\
%Keywords: Fermionic zero modes, General Dirac equation, Vortex
%background.}
\maketitle

\section{Introduction}
The study of gravity, as described by Einstein's General Relativity (GR), has been profoundly successful in explaining various phenomena in the universe. However, certain challenges arise when considering its applicability in extreme conditions, such as those found in the  black hole singularities \cite{Penrose:1964wq}, and the current accelerated expansion of the universe \cite{Caldwell:2009ix}. In particular, the enigmatic nature of dark energy \cite{Copeland:2006wr} and dark matter \cite{Garrett:2010hd} poses significant questions that GR alone cannot adequately address.
To tackle these issues, modified gravity theories have emerged as a compelling avenue of research. These theories aim to extend or amend the framework of GR, providing alternative descriptions of gravitational interactions. For instance, Brans-Dicke theory \cite{Brans:1961sx} incorporates a scalar field that varies the gravitational coupling constant, proposing a more flexible approach to understanding gravitational interactions. The ghost-free massive gravity model, known as the de Rham-Gabadadze-Tolley (dRGT) theory \cite{deRham:2010ik,deRham:2010kj}, introduces a massive graviton, solving the instability issues in earlier models and offering potential explanations for dark energy and the universe's accelerated expansion.

Higher-order gravity theories represent another significant modification, wherein higher-order derivatives of the metric are included in the gravitational field equations. These theories, such as the Einstein-Gauss-Bonnet theory, not only yield new solutions but also potentially resolve singularities that appear in classical GR. The addition of Gauss-Bonnet terms is particularly relevant in higher-dimensional spacetimes, where they can influence the dynamics of cosmological evolution. Similarly, f(R) gravity introduces a function of the Ricci scalar R into the gravitational action \cite{DeFelice:2010aj}, allowing for dynamic cosmological models that can account for the observed acceleration of the universe's expansion \cite{Carroll:2003wy}. Over the past few decades, higher-order gravity theories have garnered significant attention. On one hand, incorporating higher-curvature interactions into the Einstein-Hilbert action can yield a renormalizable theory of gravity \cite{Stelle:1976gc,Stelle:1977ry}. On the other hand, within the framework of string theory in the low-energy limit, it is generally expected that the effective action will include a series of higher-derivative terms involving the Riemann tensor and its covariant derivatives \cite{Gross:1986mw,Green:2003an,Frolov:2001xr}. These corrections not only offer new insights into gravitational behavior but also pave the way for exploring the quantum nature of gravity.

Higher-order gravity theories often face a critical issue: the existence of ghost fields. To address this problem, it is particularly important to seek ghost-free theories. 
Einsteinian cubic gravity (ECG) \cite{Bueno:2016xff} is a theory constructed from up to cubic powers of the Riemann tensor. Apart from the absence of ghost fields, its spectrum is the same as that of General Relativity. Additionally, the coefficients of each cubic invariant are dimension-independent, and notably, in four dimensions, the theory is neither trivial nor topological. For higher-order gravitational theories, studying their stationary solutions is one of the key aspects. Consequently, significant work has been conducted to explore black hole solutions. For example, Refs. \cite{Hennigar:2016gkm, Bueno:2016lrh} first presented the spherically symmetric black hole solutions, while Refs. \cite{Burger:2019wkq,Adair:2020vso} considered the rotating case.

In this paper,  we reinvestigate the stationary spherically symmetric numerical solutions of Einsteinian cubic gravity. In addition to the well-known spherically symmetric black hole solutions, we identify a new class of two-parameter solutions without event horizon. These solutions exhibit a curvature divergence at the center, categorizing them as naked singularities. Our analysis reveals that under certain conditions, the metric function can asymptotically approach zero, signaling the formation of a critical event horizon. This phenomenon has also been observed in previous studies \cite{Wang:2023tdz,Yue:2023sep,Huang:2023fnt,Ma:2024olw,Huang:2024rbg,Chen:2024bfj}. Unlike earlier research that required coupling with matter fields, in our case, with only pure gravity, this object can be described as a frozen
gravitational stars.

The paper is organized as follows.  In Section. \ref{model}, We provide an introduction to the model of Einsteinian cubic gravity, and analyze the boundary conditions for two types of solutions: black holes and those without event horizon.
 In Section \ref{sec3}, we present numerical results of those without event horizon and a comprehensive analysis of their physical properties. The conclusion and discussion
are given in the last section.

\section{EINSTEINIAN CUBIC GRAVITY}\label{model}
In this section, we provide a brief overview of Einsteinian cubic gravity. This theory acts as a higher-curvature extension of Einstein gravity (up to cubic order), comprising cubic products of the Riemann tensor 
$R_{\mu \rho \nu \sigma}$, the Ricci tensor $R_{\mu \nu }$, and the Ricci scalar $R$.
The expression for the bulk action is written as follows:
\begin{equation}\label{action}
  S=\frac{1}{16 \pi G} \int\sqrt{-g}\,d^4x\left(R- G^2 \lambda   \mathcal{P}\right),
\end{equation}
where $G$ is the Newton gravitational constant, $R$ represents the Ricci scalar,  and 
$\lambda$ is a gravitational coupling with dimensions of
 length.
The density  of cubic curvature corrections
 to the Einstein-Hilbert action read
%\begin{equation}
%{\cal P} = 12 \tensor{R}{_\mu ^\rho _\nu ^\sigma} \tensor{R}{_\rho ^\gamma _\sigma ^\delta}\tensor{R}{_\gamma ^\mu _\delta ^\nu} + R_{\mu\nu}^{\rho\sigma} R_{\rho\sigma}^{\gamma\delta}R_{\gamma\delta}^{\mu\nu} - 12 R_{\mu\nu\rho\lambda}R^{\mu \rho}R^{\nu\sigma} + 8 R_\mu^\nu R_\nu ^\rho R_\rho ^\mu.
%%\end{equation}

\begin{equation}
{\cal P} = 12 \tensor{R}{_a ^c _b ^d} \tensor{R}{_c ^e _d ^f}\tensor{R}{_e ^a _f ^b} + R_{ab}^{cd} R_{cd}^{ef}R_{ef}^{ab} - 12 R_{abcg}R^{a c}R^{bd} + 8 R_a^b R_b ^c R_c ^a.
\end{equation}
In the above equation, the coefficients of each term must be precisely specified to ensure that the spectrum includes a transverse and massless graviton, similar to that in general relativity on a maximally symmetric background.

The  equation of motion, obtained by varying the action with respect to the metric $g_{\mu\nu}$,  can be  written in the form:
\begin{equation}\label{eq3}
G_{ab}=\frac{\partial {\cal L}}{\partial R^{acde}}R_{b}{}^{cde} - \frac{1}{2} g_{ab} {\cal L} - 2 \nabla^c \nabla^d \frac{\partial {\cal L}}{\partial R^{acdb}} = 0,
\end{equation}
where ${\cal L}=R- G^2 \lambda   \mathcal{P}$,
and
\begin{align}\label{eq4}
 \frac{\partial {\cal L}}{\partial R^{abcd}}  =& g_{a[c}g_{b]d} - 6 G^2 \lambda \big[  \,  R _{ad} R _{bc} -  R_{ac} R _{bd} +  g_{bd} R _{a}{}^{e} \
R _{ce} -  g_{ad} R _{b}{}^{e} R_{ce}  -  g_{bc} R _{a}{}^{e} R_{de}  \nonumber\\
&+  g_{ac} R _{b}{}^{e} R_{de} 
-  g_{bd} R ^{ef} R_{aecf} +  g_{bc} R ^{ef} R _{aedf} + \
 g_{ad} R ^{ef} R _{becf} - 3 R_{a}{}^{e}{}_{d}{}^{f} R _{becf} \nonumber\\
&  - g_{ac} R ^{ef} R _{bedf} + 3 R_{a}{}^{e}{}_{c}{}^{f} R _{bedf} + \tfrac{1}{2} R_{ab}{}^{ef} R _{cdef} \big] \, .
\end{align}
 
We focus on stationary, spherically symmetric solutions. A suitable metric ansatz is therefore given by:
\begin{equation}
ds^2 = -f dt^2 + \frac{dr^2}{f} + r^2 \left( d\theta^2 + \sin^2 \theta \, d\phi^2 \right),
\end{equation}
 where  $f$ is  only a function of the radial coordinate $r$. 
 Taking the above ansatz into Eq. (\ref{eq3}), we can obtain the ordinary differential
 equations of the  component   $G_r{}^r$  of Einsteins equations as follows:
\begin{align}\label{feq6} 
\frac{1}{2 G}&  \left(\frac{rf'+f-k + \Lambda r^2}{r^2} \right) + \lambda G \left( \frac{6 f f''' }{r^3} \left(r f' - 2f + 2k \right)  + \frac{6 f f''^2}{r^2} \right. 
\nonumber\\
&+ \left. \frac{24 f f''}{r^4} \left(f -k - rf' \right) +\frac{6 f'^2}{r^4} \left(4f - k\right) - \frac{24 f f'}{r^5} \left( f- k\right)\right) = 0\ .
\end{align}
The above ordinary differential equation is a third-order equation. For higher-order equations, both analytical and numerical solutions are challenging. Fortunately, this higher-order equation can be rewritten in the form of a total differential as follows:
\begin{equation}\label{fequs}
\bigg((f-1)r+G^2 \lambda \big(4f'^3
+12\frac{f'^2}{r}-24f(f-1)\frac{f'}{r^2}
-12ff''(f'-(2(f-1)/r)\big)\bigg)'=0\, .
\end{equation}
The  derivative
 in above  equation happens to be zero, which means it can be equal to a constant. Integrating the above equation (\ref{fequs}), ones can
 leads to
\begin{equation}\label{feq8}
-(f-1)r-G^2 \lambda \bigg[4f'^3
+12\frac{f'^2}{r}-24f(f-1)\frac{f'}{r^2}
-12ff''\left(f'-\frac{2(f-1)}{r}\right)\bigg]=r_0\, .
\end{equation}
 Here the constant of integration $r_0$  be proportional to the mass of the
 solution through $r_0=2G M$ \cite{Bueno:2016lrh}.

Equations (\ref{feq6}) or (\ref{feq8})  present significant challenges in obtaining a general analytical solution, thus requiring the use of numerical methods to solve these differential equations. Before proceeding with the numerical treatment of these ordinary differential equations, it is essential to impose appropriate boundary conditions for the unknown function $f$. We consider two cases: one with a horizon and one without. The first case has already been analyzed in \cite{Hennigar:2016gkm,Bueno:2016lrh}, where the regular horizon condition reduces the number of solutions from a two-parameter family to a one-parameter one. That is, when the parameters $G$ and $\lambda$ are fixed, determining the mass parameter M alone is sufficient to fully specify the solutions of the differential equations, including both the interior and exterior solutions. This is consistent with the black hole uniqueness theorem. However, in the case without a horizon, we find that the solution is no longer determined by a single parameter, but instead forms a two-parameter family. In addition to the mass parameter, another new parameter is required.

Therefore, considering the asymptotically flat behavior of the solution, we adopt the following boundary conditions for the solutions corresponding to black holes and those without even horizon, respectively:
\begin{equation}\label{equ19}
\qquad \qquad \qquad  f(\infty) = 1-\frac{2 G M}{r},  \qquad \qquad \qquad  with \;\;event\;\; horizon
\end{equation}
and
\begin{equation}\label{equ20}
\qquad \qquad \qquad  f(0) = f_0, \qquad f(\infty) = 1-\frac{2 G M}{r}, \qquad  without\;\; event\;\; horizon
\end{equation}
$f_0$  represents the value of the function 
$f$ at the center. Considering that there is no event horizon, 
$f_0$
  should be greater than zero. Subsequent numerical results show that $0< f_0 < 1$.

 Before solving these equations numerically, it is important to note that there exists a special trivial solution. Specifically, when  $\lambda=0$, the solution is a Schwarzschild black hole, which can be written in the following form:
\begin{equation}
ds^2 = -(1-\frac{2G M}{r}) dt^2 + \frac{1}{1-\frac{2G M}{r}} dr^2 + r^2 ( d \theta^2 + \sin^2\theta d \phi^2).
\end{equation}
In the subsequent numerical results, we will see that even when $\lambda\neq 0$, the solutions without event horizon in certain extreme cases remain related to the Schwarzschild black holes.

\section{Numerical results}\label{sec3}
In this section, 
we will solve the Eq. (\ref{feq8}) with boundary conditions (\ref{equ19}) numerically.  For simplicity, we choose $G=1$.
To facilitate the calculations, we redefine the radial coordinate 
$r$  by introducing a new variable 
$x=\frac{r}{1+r}$.
  This transformation effectively places the computational boundaries at 
 $x=0$ and $x=1$. Throughout the study, the relative error of our numerical solutions has been maintained below $10^{-8}$.

\begin{figure}[]
  \begin{center}
  \includegraphics[width=9.0cm]{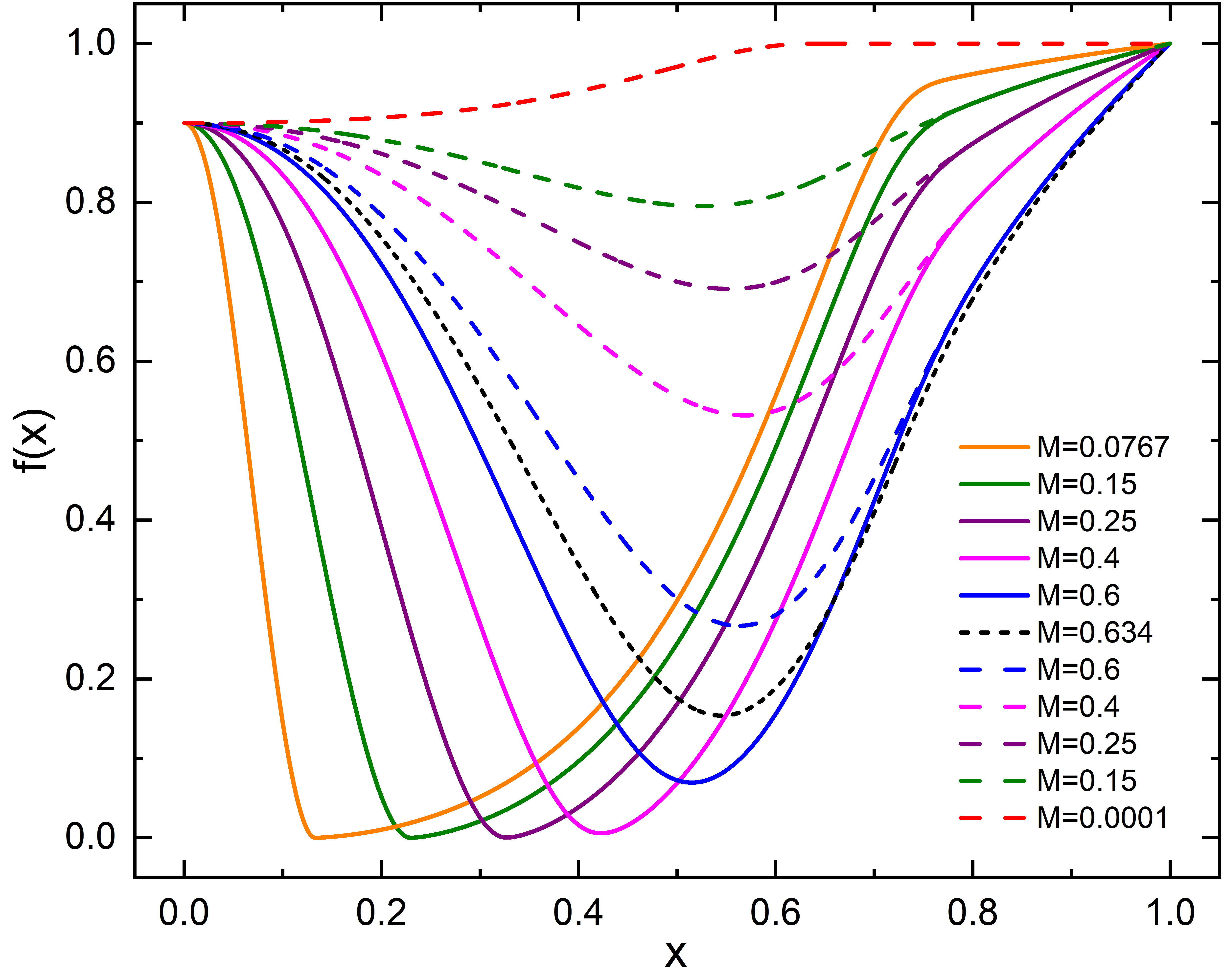}
  \end{center}
  \caption{The numerical profiles for the    function $f$ with fixed 
 parameters $f(0)=0.9$, $\lambda=0.1$. The dashed and solid lines represent two distinct branches of numerical solutions, with the same color indicating solutions of the same mass, and 
the black dotted line represents the transition between the two types of solutions.}\label{phase1}
\end{figure}
\begin{figure}[]
  \begin{center}
   \includegraphics[width=8.04cm]{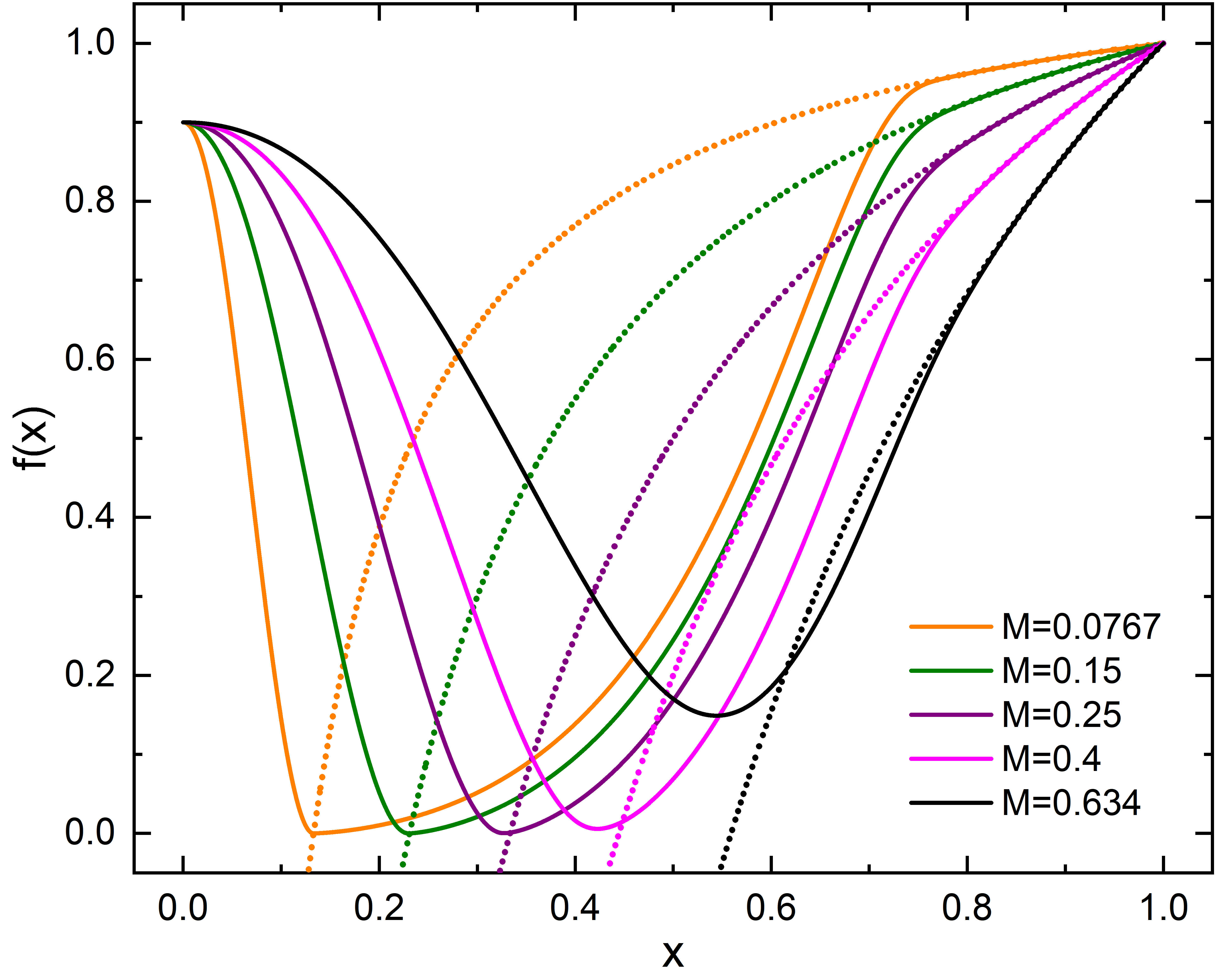}
   \includegraphics[width=8.28cm]{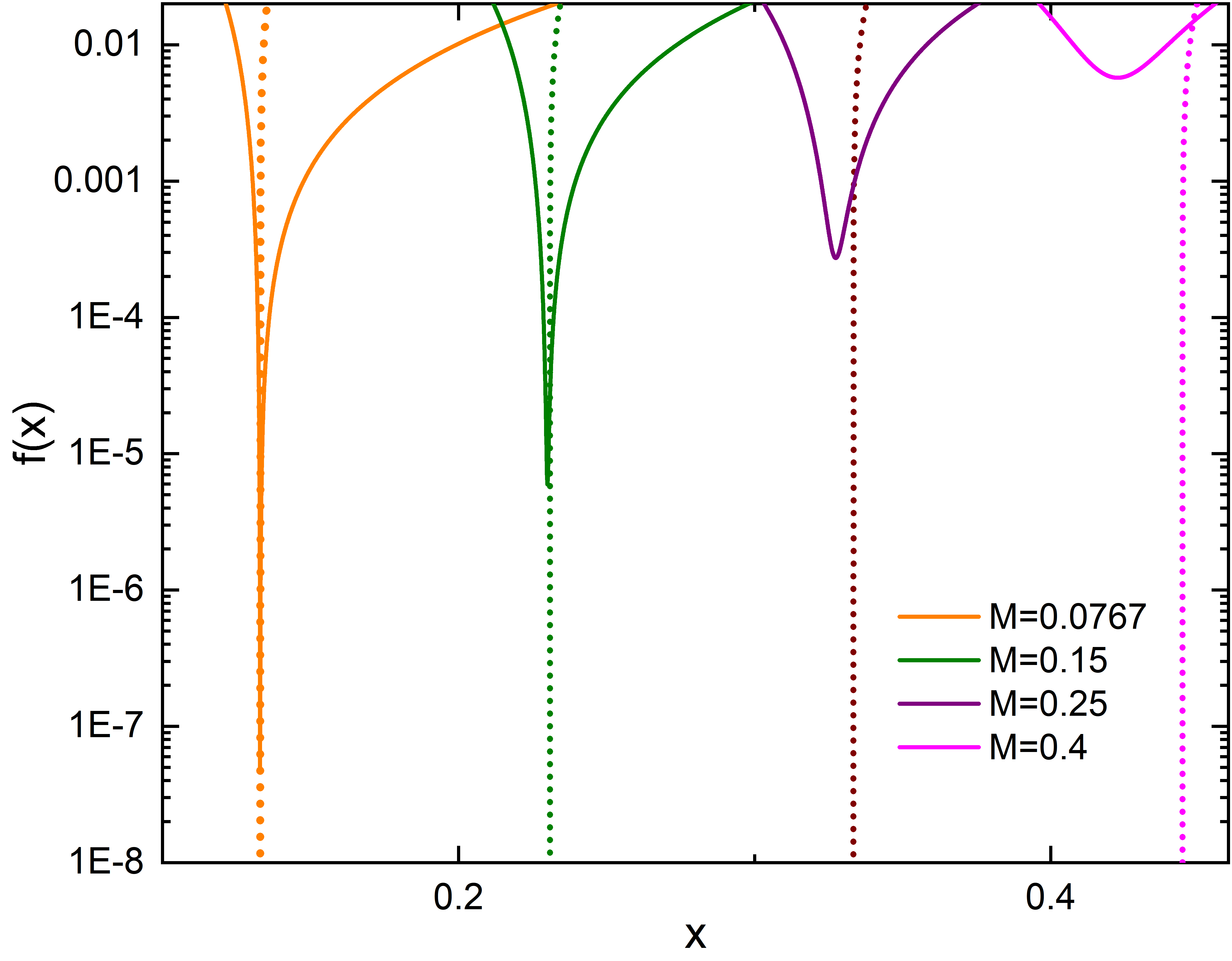}
  \end{center}
  \caption{The numerical profiles for the    function $f$ with fixed 
 parameters $f(0)=0.9$, $\lambda=0.1$. The solid lines represent the numerical solutions without an event horizon, while the dotted lines depict the Schwarzschild black hole solutions. The same color indicates solutions with the same mass.}\label{phase2}
\end{figure}

First, in Figure \ref{phase1}, we present the spatial distribution of the function $f$ with fixed parameters $f(0)=0.9$ and $\lambda=0.1$. In the case of different mass parameters $M$, we observe two distinct branches of numerical solutions, represented by dashed and solid lines, with the same color indicating solutions of same mass. The dashed branch corresponds to masses in the range \(0 < m < 0.634\), while the solid branch covers the range \(0.0767 < m < 0.634\). The transition occurs at at \(m = 0.634\), marked by a black dotted line. In the case of the dashed line, as the mass increases, the spatial curvature of \(f\) becomes more pronounced, approaching the $x$-axis. In contrast, for the solid line, as the mass decreases, \(f\)  approaches the $x$-axis, with the minimum value of \(f\) getting closer to the $x$-axis. We will discuss the details of these behaviors further in Fig. \ref{phase2}.
Notably, when \(M\) approaches zero, in standard Einstein gravity, the entire region would typically resemble Minkowski space. However, in ECG gravity, since the value of \(f(0)\) is independent of \(M\), it can remain nonzero as \(M \rightarrow 0\), indicating that the spacetime is not globally flat as \(M \rightarrow 0\). Furthermore, by calculating the  Kretschmann  scalar curvature \cite{Bueno:2016lrh}, 
\begin{equation}
R_{abcd}R^{abcd} = \frac{4(f(0) - 1)^2}{r^4} + \mathcal{O}\left(\frac{1}{r^2}\right).
\end{equation}
Given that  the constant value
$f(0)=0.9$ in Fig. \ref{phase1}, we can conclude that this type of solution essentially represents a naked singularity.

To have a clearer understanding of the behavior of the solid line branch as the minimum value of \(f\) approaches the $x$-axis, we present additional details in Fig. \ref{phase2}. The solid line represents solutions without event horizon, while the dotted line represents the Schwarzschild black hole solution of the same mass. From the plot, we observe that, due to the same mass, as \(x \to 1\) (approaching the asymptotic boundary), the solid and dotted lines coincide, which is consistent with the asymptotic condition (\ref{equ20}). In the region where \(x\) is relatively small, we find that as the mass decreases, the minimum value of \(f\) approaches zero, while the horizon position of the Schwarzschild black hole moves increasingly closer to the minimum value of \(f\). When \(M\) reaches its minimum value of \(M = 0.0767\), the position of the Schwarzschild black hole horizon nearly coincides with the minimum value of \(f\). We designate this point as a critical location, as it is situated near the threshold for event horizon formation, and thus, it can be referred to as the critical horizon.
These solutions with critical horizons, while having almost the same event horizon as Schwarzschild black holes of the same mass, appear more like an extreme black hole from an external perspective. Furthermore, with the formation of the critical horizon, the previously exposed central naked singularity can be regarded as being enveloped by the event horizon, effectively rendering the naked singularity nearly invisible. Since these solutions arise entirely from pure gravity, we refer to them as frozen gravitational stars.
It is noteworthy that, as shown in the right panel of Fig. \ref{phase2}, the minimum value of \(f\) has dropped below \(10^{-7}\). At this point, the relative error has begun to increase, and the program has crashed. It remains unclear whether the minimum mass value can go below its current level.

\begin{figure}[]
  \begin{center}
   \includegraphics[width=8.15cm]{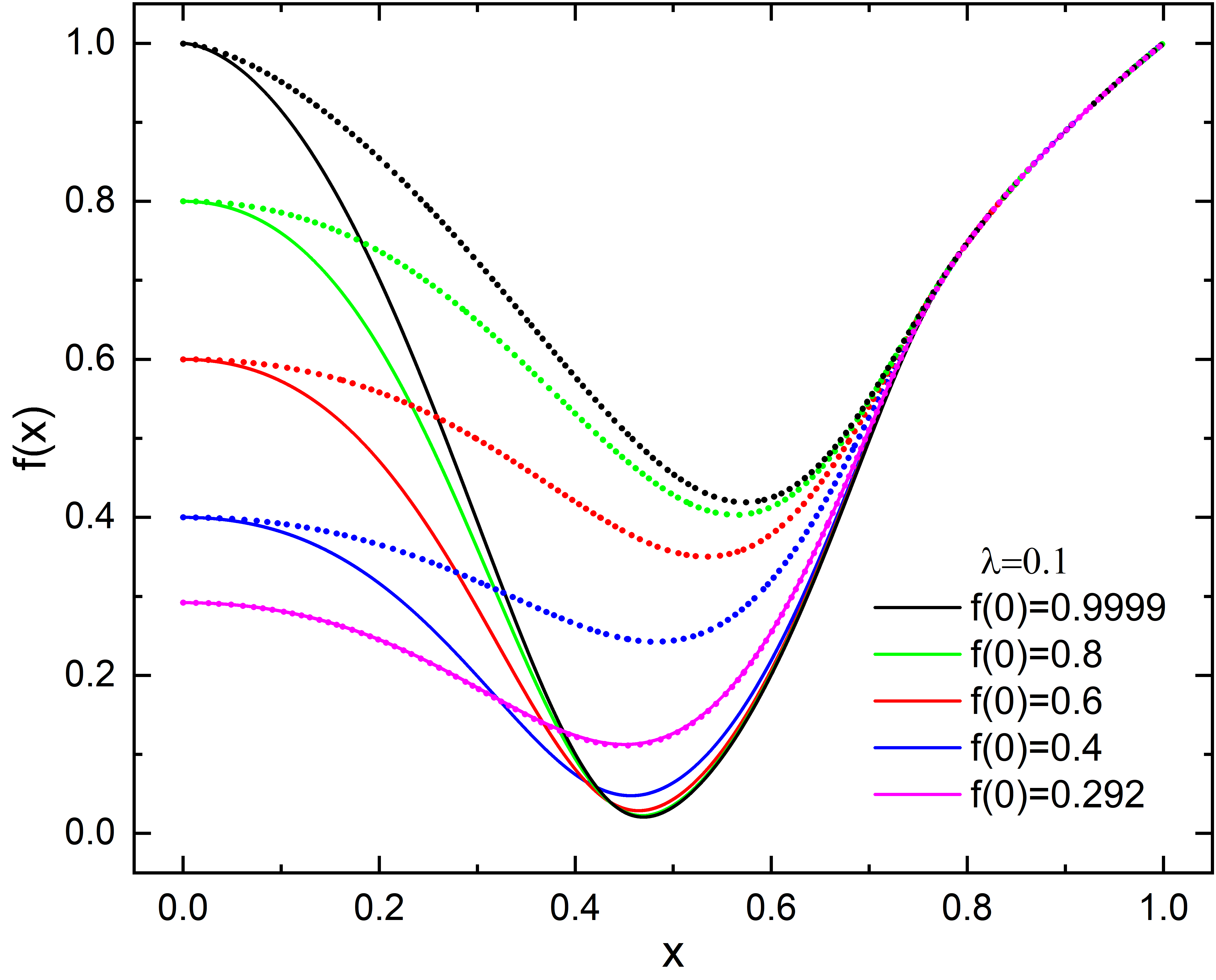}
     \includegraphics[width=8.15cm]{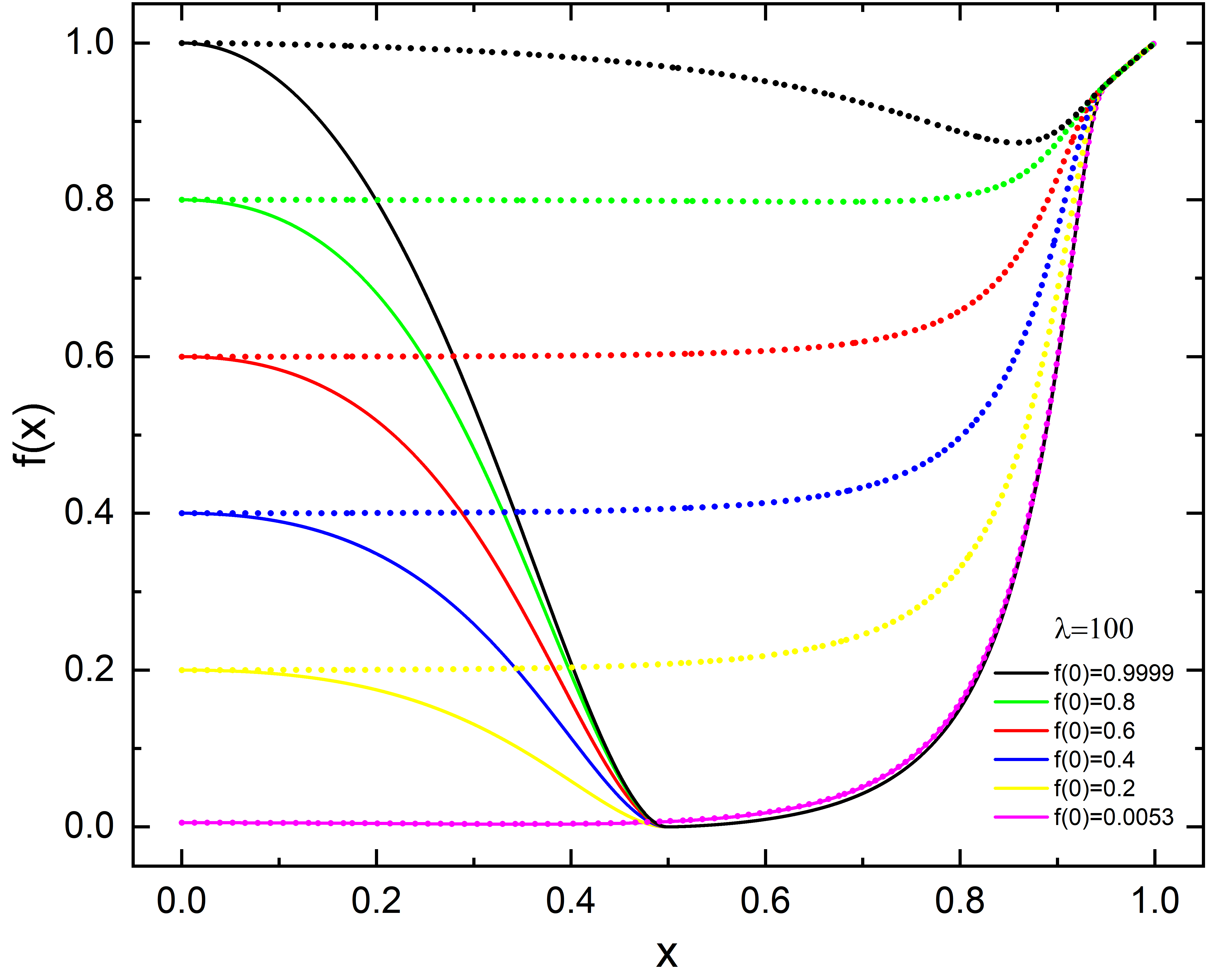}
  \end{center}
  \caption{The spatial distribution of the function $f$ with fixed parameters $M=0.5$.
 The left panel corresponds to 
$\lambda=0.1$, while the right panel corresponds to 
$\lambda=100$. The dotted and solid lines represent two distinct branches of numerical solutions, with the same color indicating solutions of the same mass. 
  }\label{phase3}
\end{figure}

\begin{figure}[]
  \begin{center}
  \includegraphics[width=8.0cm]{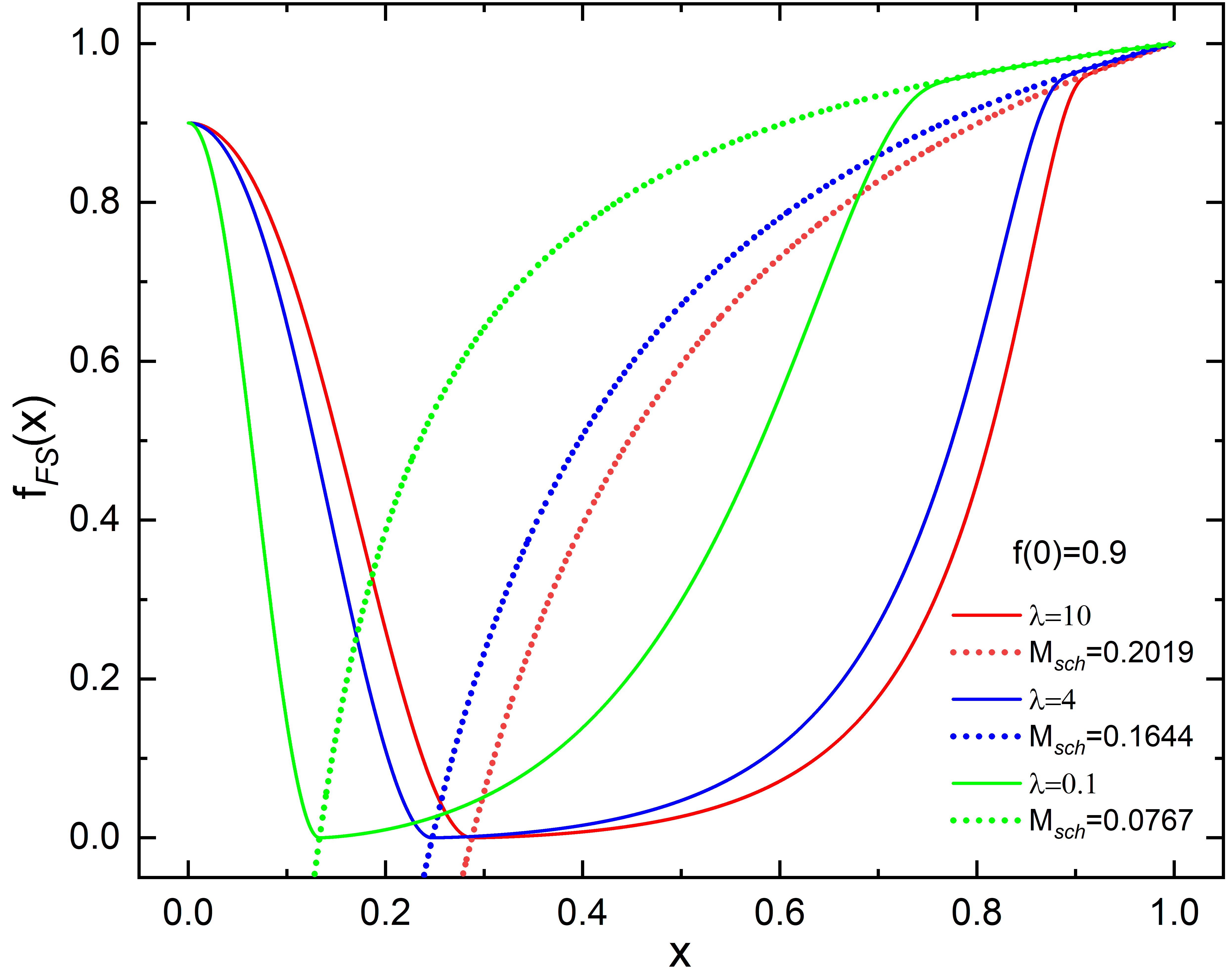}
   \includegraphics[width=8.25cm]{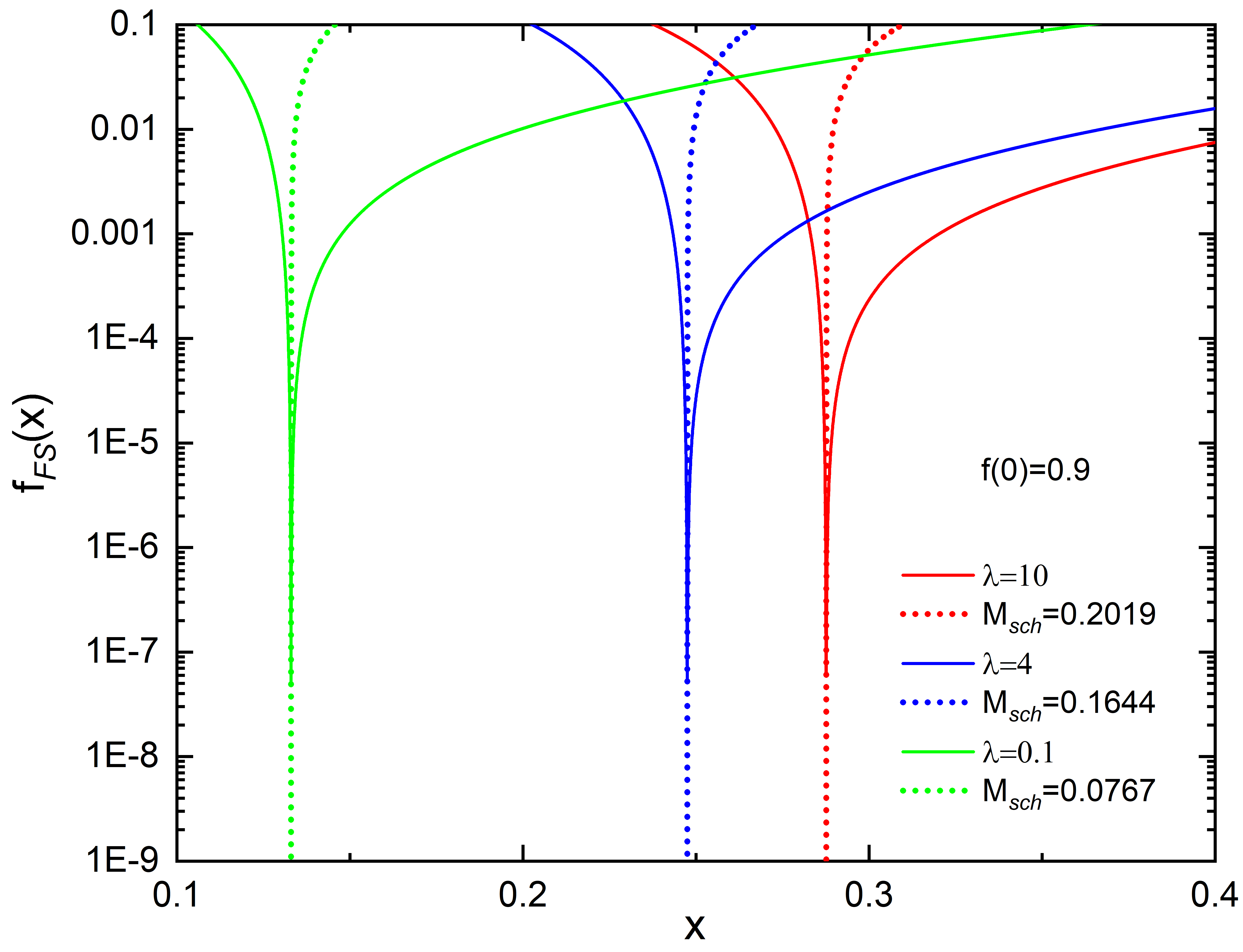}
  \end{center}
  \caption{The spatial distribution of frozen gravitational stars with varying values of 
$\lambda$ and a fixed  $f(0)=0.9$.
The solid lines represent the numerical solutions without an event horizon, while the dotted lines depict the Schwarzschild black hole solutions. The same color indicates solutions with the same mass.}\label{phase4}
\end{figure}

Since the gravitational coupling 
$\lambda$ represents the degree of deviation from Einstein's gravity, it is crucial to study how different 
$\lambda$ values affect the numerical results. 
First, in Fig. \ref{phase3}, 
 we present the spatial distribution of the function $f$ with fixed parameters $M=0.5$.
 The left panel corresponds to 
$\lambda=0.1$, while the right panel corresponds to 
$\lambda=100$.
 For different values of the parameter \(f(0)\), there are two distinct branches of numerical solutions, represented by dotted and solid lines, with the same color indicating solutions with the same \(f(0)\) value. 
 In the case of the dotted line, as the    value of \(f(0)\) decreases, the minimum value of \(f\) getting closer to the $x$-axis. In contrast, for the solid line, as the    value of \(f(0)\) decreases, decreases,  the minimum value of \(f\) increases. Furthermore, when \(\lambda\) is large, the solutions of the solid branch all approach the $x$-axis at nearly  the  same position, with the external function distribution remaining nearly identical. This suggests that when \(\lambda\) is sufficiently large, the second branch will give rise to FGSs.

In Fig. \ref{phase4}, 
We present   how the magnitude of 
$\lambda$ may influence the properties of frozen gravity stars.
The three solid lines in green ($\lambda=0.1$), blue ($\lambda=4$), and red ($\lambda=10$) represent frozen gravity stars, while the corresponding dashed lines in the same colors represent Schwarzschild black hole solutions of equal mass. From the right panel, we can clearly see that regardless of the value of 
$\lambda$, the critical horizon position of the FGSs is almost identical to the event horizon position of the corresponding Schwarzschild black holes. Moreover, as 
$\lambda$ increases, this critical position also rises.

\begin{figure}[]
  \begin{center}
  \includegraphics[width=8.1cm]{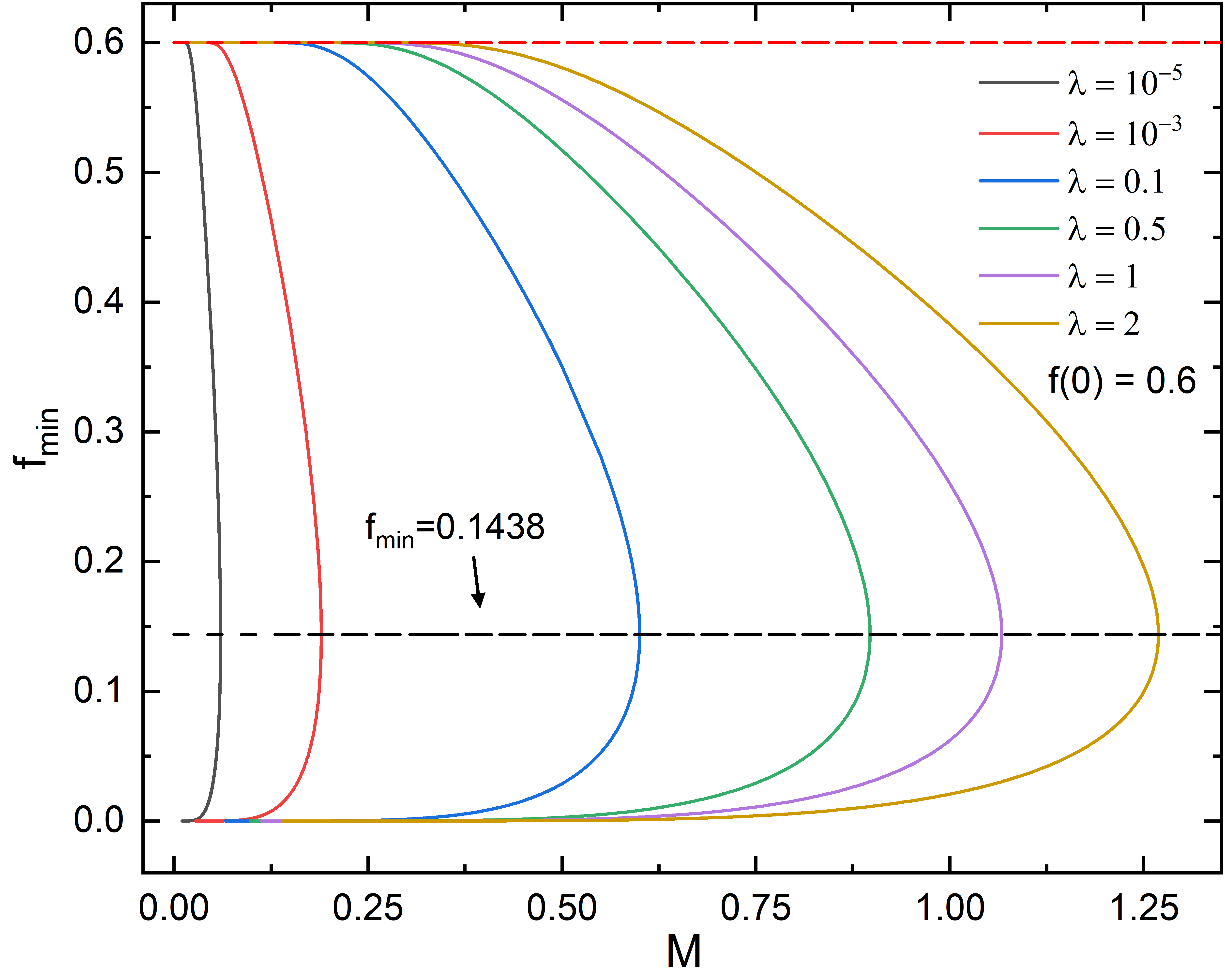}
   \includegraphics[width=8.1cm]{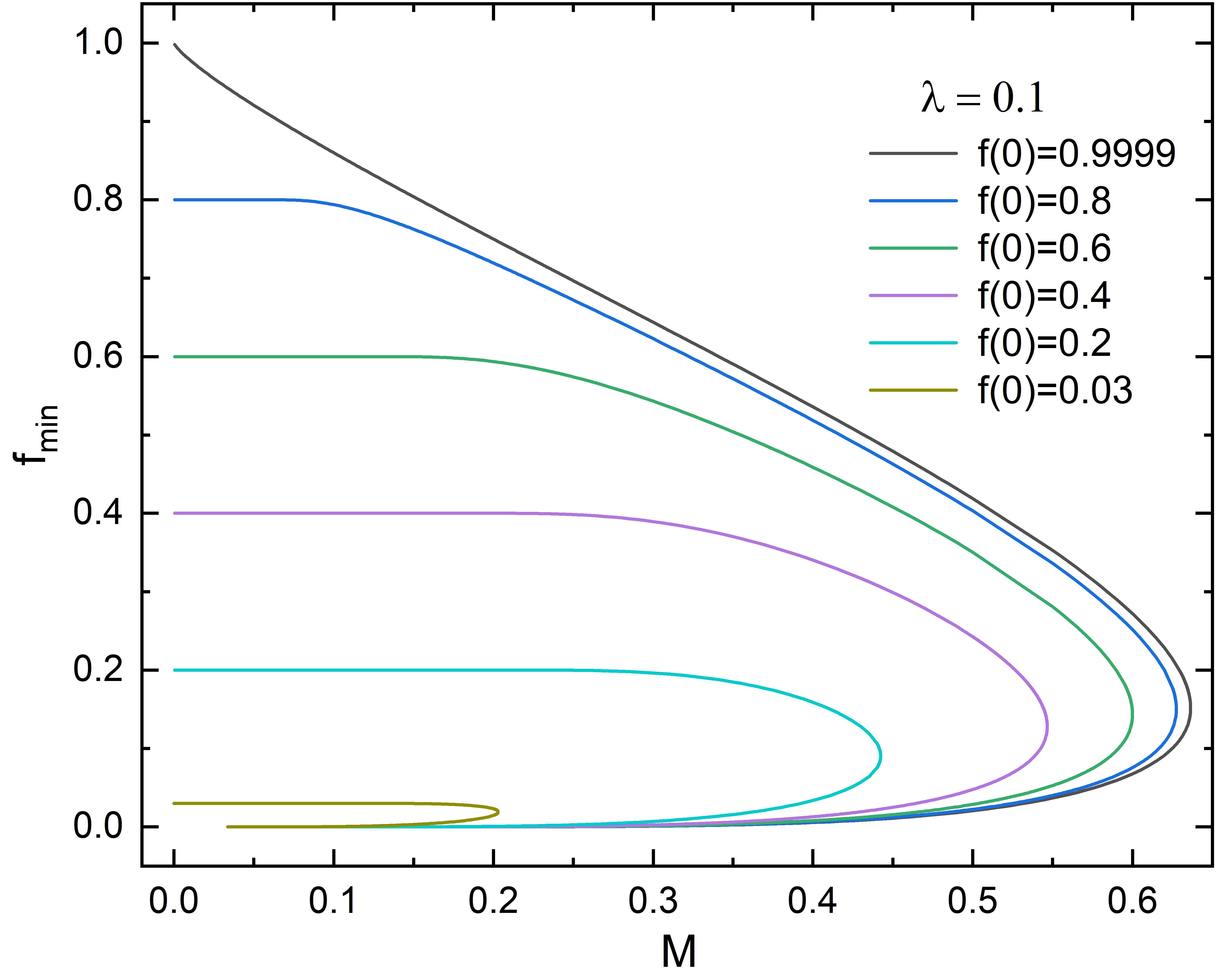}
  \end{center}
  \caption{The minimum value of the function \(f\) versus the mass $M$ with the different values of $\lambda$ (Left panel), and with the different values of $f(0)$ (Right panel).  }\label{phase5}
\end{figure}

To have a comprehensive understanding of the branch of solutions without a horizon, we present in Fig. \ref{phase5} the relationship between the minimum value of the function \(f\) and the mass \(m\). The left and right panels correspond to fixed parameters \(m\) and \(\lambda\), respectively. 
In the left panel, the curves of different colors represent different values of \(\lambda\), increasing from left to right. As the value of \(\lambda\) increases, the maximum mass of the solutions also increases. The horizontal red dashed line represents \(f_{\text{min}} = 0.6\), while the horizontal black dashed line represents \(f_{\text{min}} = 0.1438\). Interestingly, we find that the curves corresponding to different values of \(\lambda\) exhibit the same minimum \(f\) value at the transition point between the two branches.
In the right panel, the curves of different colors represent different values of f(0), decreasing from up to down. As the value of \(f(0)\) increases, the maximum mass of the solutions also increases. It is noteworthy that the maximum value of \(f(0)\) cannot exceed 1.

\section{Conclusions}
In this paper, we have reinvestigated the stationary spherically symmetric numerical solutions of Einsteinian cubic gravity. 
 Beyond the well-established spherically symmetric black hole solutions \cite{Hennigar:2016gkm, Bueno:2016lrh}, we have identified a new kind of solutions that lack an event horizon. These solutions require an additional parameter, making them two-parameter solutions, in contrast to black hole solutions that are characterized by a single parameter $M$.
Significantly, these new solutions exhibit a divergence in curvature at the center, effectively categorizing them as naked singularities. Our analysis of the solutions reveals that, for any non-negligible value of the gravitational coupling parameter $\lambda$, there are consistently two distinct branches of solutions. One branch degenerates to the case of $M \to 0$, which does not correspond to the trivial Minkowski spacetime. In contrast, the other branch leads to the emergence of a critical horizon when the mass approaches a specific limit, transitioning into a frozen gravity star. At this critical horizon, the component $g_{tt}$ approaches zero, causing the  frozen gravitational star  to exhibit properties akin to an extreme black hole from an external perspective. Consequently, due to the emergence of critical horizon, this configuration obscures the internal naked singularity.

Future investigations can be expanded in the following areas. Firstly, the solutions we found within the framework of Einsteinian cubic gravity theory represent a class of naked singularities. These singularities can be addressed by introducing matter fields, such as scalar fields, to eliminate the singularity and obtain solutions akin to boson stars \cite{Liebling:2012fv}. Secondly, the current research can be extended to Anti-de Sitter  or de Sitter  spacetimes to study the impact of the cosmological constant on the properties of frozen gravity stars.
 Finally, the stability of these new sulutions remain unexplored in this paper, and we intend to address this aspect in our further work.

\section{Acknowledgment}
This work is supported by National Key Research and Development Program of China (Grant No. 2020YFC2201503) and the National Natural Science Foundation of China (Grants No.~12275110 and No.~12047501).

\end{document}